# Compressive Circular Polarization Snapshot Spectral Imaging

JIANGLAN NING[1], ZHILONG XU[1], DAN WU[2], RUI ZHANG[1], YUANYUAN WANG[1], YINGGE XIE[1], WEI ZHAO[3], XU MA[4], AND WENYI REN[1,*]

[1]School of Science, Northwest A&F University, Yangling, Shaanxi 712100, P. R. China
[2]College of Mechanical and Electronic Engineering, Northwest A&F University, Yangling, Shaanxi 712100, P. R. China
[3]Shaanxi Public Security Judicial Authentication Center, Shaanxi Public Security Department, Xi'an, Shaanxi 710016, P. R. China
[4]Key Laboratory of Photoelectronic Imaging Technology and System of Ministry of Education of China, School of Optoelectronics, Beijing Institute of Technology, Beijing 100081, China
*renovelhuman@gmail.com

A compressive sensing based circular polarization snapshot spectral imaging system is proposed in this paper to acquire two-dimensional spatial, one-dimensional circular polarization (the right and left circular polarization), and one-dimensional spectral information by one snapshot measurement, simultaneously. The dispersed component combined with a double Amici prism and a Wollaston prism was used to implement the spectral shifting along two orthogonal directions. Theoretically, the spectral resolution could be improved compared with the traditional coded aperture snapshot spectral imager proposed by Brady et al. The right and left circular polarization components are extracted by the assembly with an achromatic quarter wave-plate and a Wollaston prism. The encoding and reconstruction are illustrated comprehensively. The feasibility and fidelity are verified from the aspects of image and spectral reconstruction evaluations. It provides us an alternative approach for circular polarization spectral imaging for the dynamic objects in the fields such as defogging, underwater imaging, insects detection, and so on.

## 1. INTRODUCTION

Intensity, spectrum, and polarization information as the most fundamental three properties of light, can be taken by the camera, spectrometer, and polarimeter, respectively. The structure, molecular, and functional information of materials can be identified based on spectral feature [1]. The roughness, edge, and physical information can be recognized via the polarization information. Imaging spectropolarimetry (ISP) as a promising multidimensional imaging technology, can detect the spatial, spectral, and polarization information of targets, simultaneously. According to the amount of information taken in polarization domain, ISP can be categorized as the linear, circular, and full-Stokes polarization imaging systems [2]. Due to the out performance of circular polarization in maintaining the intended polarization state for large optical depths, it is important in many critical sensing applications such as underwater imaging, defogging imaging, and so on [2, 3]. Some imagers were proposed to detect the objects in the turbid medium using the circular polarization information [4-6]. However, the spectral information cannot be acquired. For other applications, some full Stokes ISPs were developed by combing various imaging spectrometer and polarimeter [7-11]. These systems are designed based on the Nyquist-Shannon sampling theorem and puzzled by the problems such as requiring scanning in different domains to implement high-dimensional imaging, the valuable large format sensor for infrared detection, the huge amount of data to transfer in remote sensing, the harm from high doses X-ray radiation in biomedical imaging, and so on.

Compressive sensing provides us with a powerful tool to recover the signal from far fewer samples than required by the Nyquist-Shannon sampling theorem and are widely applied in numerous fields such as imaging, communication, facial recognition, and so on [12]. Thereby, the abovementioned problems can be mitigated based on the compressive sensing framework. Two typical cases of compressed sensing used in imaging are single pixel camera [13] and coded aperture snapshot spectral imager (CASSI) [14,15]. Combining the compressive sensing based spectral imaging and polarimetry, several compressive ISPs were proposed in the last years. A compressive linear polarization spectral imager was proposed by Tsung-Han Tsai al. based on the uniaxial crystal and CASSI in 2013 [16]. A compressed spectral polarization imager driven by a rotating prism and a color detector with a pixelated polarizer array (PPA) was developed by Chen et al. in 2014 [17]. PPA is used as a polarization encoder. In 2015, a compressed spectral polarization imager based on a pixelated polarizer and a color pattern detector was proposed by Chen et al. [18]. A linear polarization spectral imager was developed by Yuan et al. based on liquid crystal modulator in 2015 [19]. The spatial distribution of four linear polarizations and three-color channels was captured. The polarization encoding was exerted by the spatial light modulator. In 2019, Ren et al. developed a channeled compressive full-Stokes spectropolarimetry imager by combing the channeled spectral modulator with CASSI [20, 21]. A polarizer filter wheel was used to enhance the polarization encoding. All the systems proposed above based on compressive sensing are operated by scanning. It is not suitable for the dynamic object detection [22].

In this paper, we demonstrate a spectral polarization imaging strategy, referred to compressive circular polarization snapshot spectral imager (CCPSSI) which is able to detect both the static and dynamic objects featured with circular polarization via one snapshots measurement without scanning. The paper is organized as following. Section 2 places the principle and theory of CCPSSI. Section 3 details the simulation and evaluation to verify the feasibility. The conclusion is presented in the $4^{th}$ section. The final section gives a brief acknowledgement.

## 2. PRINCIPLES AND THEORY

### 2.1. Principle of CCPSSI

The schematic of CCPSSI is shown in Fig. 1. The incident passed through the objective lens (L1) is firstly imaged and spatially coded by the binary coded aperture (CA) with the transmittance of 50%. The coded beam is collimated by lens L2 and dispersed along the vertical direction by the double Amici prism (DAP). Then, the vertical dispersed light is modulated by the quarter wave plate (QWP). Then, the modulated light is dispersed along the horizontal direction and split into two orthogonal polarization components (the right and circular light) by the Wollaston prism (WP) [23]. Finally, the right circular polarization(RCP) and left circular polarization(LCP) components impinge on the focal plane array (FPA) through imaging lens L3. The coded pattern finally can be acquired by the FPA. For the convergence beams, the focal lengths of the two beams from WP are different [23].

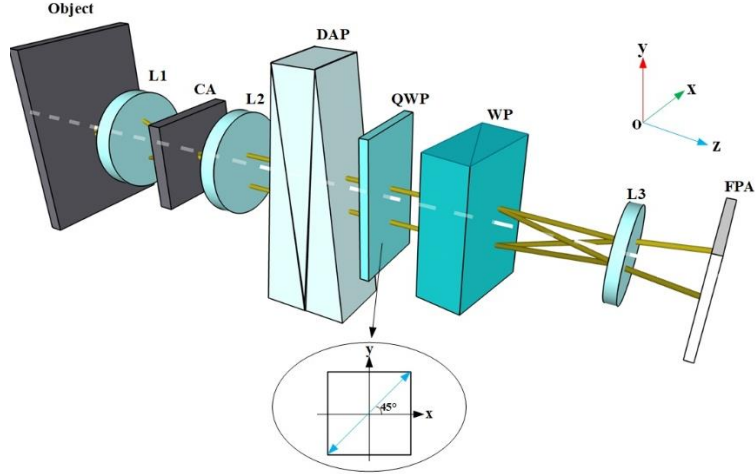

Fig. 1. Schematic of CCPSSI system

It is assumed that $\mathbf{S}(x,y,\lambda) = [S_0(x,y,\lambda), S_1(x,y,\lambda), S_2(x,y,\lambda), S_3(x,y,\lambda)]^T$ is the Stokes parameter of the object. $\lambda$ is the wavelength and $(x,y)$ are the spatial coordinates of the object. $S_0$ is the total photon flux. $S_1$ denotes preference for 0° over 90° linear polarization. $S_2$ denotes preference for 45° over 135° linear polarization. $S_3$ denotes preference for RCP over LCP. The Stokes parameter of the incident light is simplified as $\mathbf{S} = [S_0, S_1, S_2, S_3]^T$. The light passing through the QWP and WP is calculated by

$$\mathbf{S}' = \mathbf{M}_{WP} \mathbf{M}_{QWP} \mathbf{S}, \qquad (1)$$

where $\mathbf{M}_{WP}$ and $\mathbf{M}_{QWP}$ are the Mueller matrices of WP and QWP, respectively. The fast axis angle of QWP with respect to the x axis is $45°$. $\mathbf{M}_{QWP}$ is given by [24].

Wollaston prism can be regarded as a retarder or a polarization beam-splitter. As a beam-splitter, it is equivalent to a cascaded polarized component composed by the horizontal and vertical polarizers [26]. The general Muller matrix of a WP with transmission axis angle of $\beta$ to the x direction is given by

$$\mathbf{M}_{WP}(\beta) = \frac{1}{2} \begin{bmatrix} 1 & \cos 2\beta & \sin 2\beta & 0 \\ \cos 2\beta & \cos^2 2\beta & \cos 2\beta \sin 2\beta & 0 \\ \sin 2\beta & \cos 2\beta \sin 2\beta & \sin^2 2\beta & 0 \\ 0 & 0 & 0 & 0 \end{bmatrix}, \qquad (2)$$

where $\beta=0°$ or $90°$ for our system. which means that the two transmission axes of the WP are defined in the x and y directions, respectively. Substitute $\mathbf{M}_{WP}$ and $\mathbf{M}_{QWP}$ into Eq. (1), can be obtained. It can be gotten that

$$S'_R = 1/2[S_0+S_3, S_0+S_3, 0, 0]^T$$
$$S'_L = 1/2[S_0-S_3, S_0-S_3, 0, 0]^T \quad . \tag{3}$$

The subscripts 'R' and 'L' represent the RCP and LCP components. The FPA acquired RCP and LCP lights are expressed as $f'_R = 1/2(S_0+S_3)$ and $f'_L = 1/2(S_0-S_3)$ which are derived while $\beta$ is 0° and 90°, respectively. Degree of circular polarization (DoCP) and angle of circular polarization (AoCP) are two critical criteria to evaluation the significance of circular polarization light and defined as

$$\text{DoCP} = |S_3|/S_0$$
$$\text{AoCP} = \frac{1}{2}\arcsin(\frac{S_3}{S_0}) \quad . \tag{4}$$

## 2.2. Spectral Encoding

As shown in Fig. 2, the RCP and LCP spectral information are represented as two three-dimensional $(x, y, \lambda)$ hyperspectral datacubes $f_0(x, y, \lambda)$. $f_R(x, y, \lambda)$ and $f_L(x, y, \lambda)$ represent the RCP and LCP information about objects, respectively. The subscripts 'L' and 'R' in this paper denote the RCP and LCP. Let $T(x, y)$ be the transmittance function of the binary CA. The bright and dark pixels denote the transparent and opaque parts, respectively. The beam firstly hits on the binary CA and is encoded in the spatial domain $(x, y)$ as datacubes $f_1(x, y, \lambda)$. The spatially encoded datacubes $f'_1(x, y, \lambda)$ through the DAP are dispersed along x-axis. The spatial encoded and sheared datacube is transformed into the RCP and LCP components by the module combined with QWP and WP. The datacubes $f_2(x, y, \lambda)$ with respect to the RCP and LCP components extracted by WP are sheared along y-axis toward the opposite directions due to the dispersion of WP. Finally, the FPA records the projected pattern.

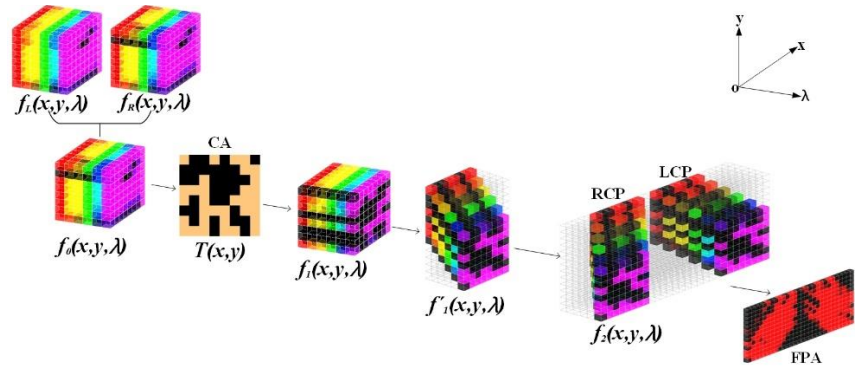

Fig. 2. The encoding flowchart of CCPSSI.

The complete information of the object can be given as

$$f_0(x, y, \lambda) = f_R(x, y, \lambda) + f_L(x, y, \lambda) \tag{5}$$

$f_0(x, y, \lambda)$ is spatially encoded by the binary CA. The modulated information is described as

$$f_1(x, y, \lambda) = T(x, y) f_0(x, y, \lambda) . \tag{6}$$

$f_1(x, y, \lambda)$ is dispersed along the vertical direction (y axis) by the DAP. The vertical shifted information $f_1'(x, y, \lambda)$ can be described by

$$f_1'(x, y, \lambda) = \iint \delta(x' - x) \delta(y' - (y + S_1(\lambda))) f_1(x', y', \lambda) dx' dy' , \tag{7}$$

where $S_1(\lambda) = \alpha_1(\lambda)(\lambda - \lambda_c)$, $\alpha_1(\lambda)$ is the wavelength dependent dispersion coefficient induced by the DAP. $\lambda_c$ is the central wavelength.

RCP and LCP information would be extracted out while the light passed through the QWP and WP. Meanwhile, the symmetric horizontal shift is introduced by the WP. The extracted shifted beams are formulated as

$$\begin{aligned} f_2(x, y, \lambda) &= \iint \delta(x' - (x + S_2(\lambda))) \delta(y' - (y + S_1(\lambda))) f_1(x', y', \lambda) dx' dy' \\ &= f_1(x + S_2(\lambda), y + S_1(\lambda), \lambda) T(x + S_2(\lambda), y + S_1(\lambda)) \end{aligned}, \tag{8}$$

where, $S_2(\lambda) = \pm \alpha_2(\lambda)(\lambda - \lambda_c)$. $\alpha_2(\lambda)$ is the wavelength dependent dispersion coefficient of the WP. The symbol "$\pm$" denote the RCP and LCP are shifted along the opposite directions along the x axis. The integrating of the image on the FPA within the spectral range $\Lambda$ is

$$g(x, y) = \int_\Lambda \omega(\lambda) f_1(x + S_2(\lambda), y + S_1(\lambda), \lambda) T(x + S_2(\lambda), y + S_1(\lambda)) d\lambda , \tag{9}$$

where $\omega(\lambda)$ is the spectral response function of FPA.

FPA is spatial pixelated. $g(x, y)$ is sampled and integrated across the grid of the detector. The measurement at the $(m, n)^{th}$ pixel is

$$g_{mn} = \iint D(m, n, x, y) g(x, y) dx dy + \eta_{mn} , \tag{10}$$

where $\eta_{mn}$ is additive noise, and $D(m, n, x, y)$ is given by

$$D(m, n, x, y) = rect(x/\Delta - n) \otimes rect(y/\Delta - m) ,$$

where $\otimes$ and rect represents the Kronecker product and rectangle function, respectively. $\Delta$ is the pixel pitch.

### 2.3. Model discretization

It is assumed that $t(i, j)$ represent the value of the $(i, j)^{th}$ pixel on the pixelated CA. The transmittance function $T(x, y)$ can be discretized as

$$T(x,y) = \sum_{i,j} t(x,y)\tau(i,j,x,y) , \tag{11}$$

where $q$ is an integer, and $\tau(i,j,x,y) = rect(x/q\Delta - j) \otimes rect(y/q\Delta - i)$.

Substituting Eq. (10) into Eq. (8), it is can be obtained that

$$g(x,y) = \sum_{i,j} t(i,j) \int_\Lambda \omega(\lambda) f_1(x+S_2(\lambda), y+S_1(\lambda), \lambda) \tau(i,j,x+S_2(\lambda), y+S_1(\lambda)) d\lambda. \tag{12}$$

To achieve the spectral discretization, the spectral range could be divided into a limited number of subintervals $\Lambda_k$. Then, Eq. (12) is modified as

$$\begin{aligned} g(x,y) &= \sum_{i,j,k} t(i,j) \int_{\Lambda_k} \omega(\lambda) f_1(x+S_2(\lambda), y+S_1(\lambda), \lambda) \tau(i,j,x+S_2(\lambda), y+S_1(\lambda)) d\lambda \\ &= \sum_{i,j,k} t(i,j) \omega_k f_1(x+s_{2k}, y+s_{1k}, \lambda_k) \tau(i,j,x+s_{2k}, y+s_{1k}) \upsilon_k + \tilde{\eta}_k \end{aligned}, \tag{13}$$

Where, $s_{1k} = S_1(\lambda_k)$, $s_{2k} = S_2(\lambda_k)$. $\lambda_k$ is in the subinterval $\Lambda_k$. $\upsilon_k$ is quadrature weight. $\omega_k$ is the spectral response of FPA with respect to the k$^{th}$ wavelength (subinterval), and $\tilde{\eta}_k$ is the discretization additive noise. Equation (13) can be rewritten as

$$g(x,y) = \sum_{i,j,k} t(i-s_{ik}, j-s_{jk}) \tau(i-s_{ik}, j-s_{jk}, x+r_{jk}, y+r_{ik}) \omega_k f_1(x,y,\lambda_k) \upsilon_k + \tilde{\eta}_k , \tag{14}$$

where $i - s_{ik}$ and $j - s_{jk}$ are the CA overlapping with $\Lambda_k$, respectively. Then let's $r_{jk} = s_{1k} - s_{jk}$ and $r_{ik} = s_{2k} - s_{ik}$. Substitute Eq. (14) into Eq. (10), $g_{mn}$ can be given by

$$g_{mn} = \sum_{i,j,k} h(\{m,n,i,j,k\}) f(i,j,k) + \eta_{ijk} , \tag{15}$$

where, $f(i,j,k) = \omega_k \upsilon_k f_1(x_j, y_i, \lambda_k)$, and $h(\{m,n,i,j,k\})$ is expressed as

$$h(\{m,n,i,j,k\}) = \iint D(m,n,x,y) \sum_{s_{ik}, s_{jk}} t(i-s_{ik}, j-s_{jk}) \tau(i-s_{ik}, j-s_{jk}, x+r_{jk}, y+r_{ik}) dxdy .$$

The matrix formation of the discrete measurement equation in Eq. (15) is

$$g = \sum_k \mathbf{H_k} f_k + \eta = \mathbf{H} f + \eta , \tag{16}$$

where $\mathbf{H_k}$ and $f_k$, respectively, are given by

$$\begin{aligned} \mathbf{H_k} &= \left[ h(\{:,:\},\{:,:,k\}) \right] \\ f_k &= \left[ f(\{:,:,k\}) \right] \end{aligned}. \tag{17}$$

An example of the sensing matrix $\mathbf{H}$ is illustrated in Fig. 3. Due to the spatial dispersion of WP, LCP and RCP states are generated after bands passing through WP. This matrix has a diagonal pattern where the circled diagonal vectors repeating horizontally correspond to the distribution of the coded aperture used for both the RCP (top left) and LCP (bottom right). The first diagonal pattern of this matrix corresponds to the first band and the rest of diagonal patterns accounts for the other five bands.

As shown in Fig. 3, under the condition that the dispersion bands do not overlap, WP disperse four bands (1$^{st}$, 2$^{nd}$, 4$^{th}$ and 6$^{th}$ bands) and DAP disperses two bands (3$^{rd}$ and 5$^{th}$ bands) in RCP. The number of dispersion bands of a single dispersion prism is much less than that of two orthogonal dispersion prism. Thereby, the spectral resolution would be improved comparing with the traditional CASSI using a single dispersion component theoretically.

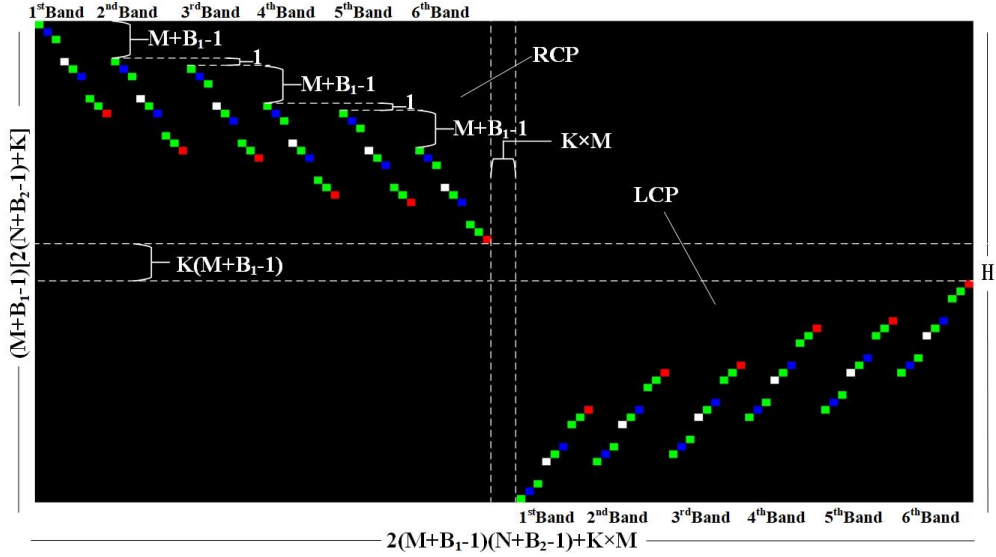

Fig. 3. Single shot sensing matrix **H** example for M=3, N=3 and L=6. $B_1$ and $B_2$ are the number of bands of vertical and horizontal sheared. RCP and LCP are coded apertures of right and left circular polarization, respectively. K denotes the gap of RCP and LCP for FPA. In sensing matrix, WP enables the circular polarization to generate horizontal dispersion, and the displacement between adjacent bands is $M+B_1-1$. The vertical dispersion generated by DAP causes the displacement between the two bands to be 1. The upper and lower distance between LCP and RCP is $K(M+B_1-1)$, and the left and right distance is $K \times M$. The size of the sensing matrix is $(M+B_1-1)[2(N+B_2-1)+K] \times [2(M+B_1-1)(N+B_2-1)+K \times M]$. The total number of wavelengths for the system is formulated as $L=B_1+B_2$. Black squares represent zero-valued elements (blocking light). Color squares represent encoding elements.

### 2.4. Reconstruction

Compressive sensing specifies that the accurate and unique solutions to underdetermined systems can be determined by solving sparsity-constrained optimization problems. There are several algorithms have been applied in the CASSI systems. For instance, the gradient projection for sparse reconstruction (GPSR), nested adaptive refinement estimation (NeARest), sparse reconstruction by separable approximation (SpaRSA), and two-step iterative shrinkage/ thresholding (TwIST) all had been used successfully [15]. In general, TwIST has produced the most accurate and visually pleasing reconstructions while being an efficient algorithm that solves the Lagrangian unconstrained formulation of constrained optimization problems [26]. Therefore, TwIST is used to reconstruct the information. The objective function is

$$\hat{f} = argmin_f \left\{ 1/2 \|g - Hf\|^2 + \tau H_{TV}(f) \right\} , \qquad (18)$$

where $\tau$ is the weighting factor of the regularization. The total variation (TV) regularizer $H_{TV}(f)$ is defined by

$$H_{TV}(f) = \sum_{l}\sum_{i,j} \left| \sqrt{\left[f(i+1,j,l)-f(i,j,l)\right]^2 + \left[f(i,j+1,l)-f(i,j,l)\right]^2} \right|, \quad (19)$$

where $i$, $j$, and $l$ denote the indexes of row, column, and wavelength, respectively.

## 3. SIMULATION AND EVALUATION

### 3.1. SIMULATION

In this paper, the data used in the simulation were acquired by a hyperspectral imaging spectropolarimeter system in our laboratory. The true color image of the objects is shown in Fig. 4. The hyperspectral dataset with 25 bands are used. The TwIST algorithm was applied to recover the information.

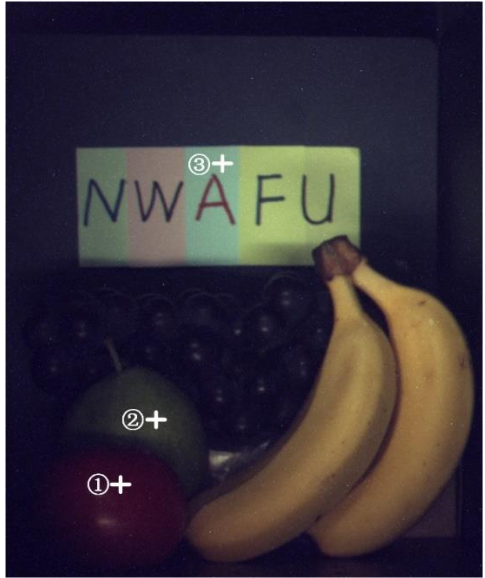

Fig. 4. The RGB image of the fruit.

Figure 5(a) shows the simulated results of the CAs with respect to the RCL at various wavelengths. In practical experiment, the CAs are taken in the calibration via illuminating the system by different monochromatic sources. The dispersed CA images alternately shift along the horizontal and vertical directions as the wavelength increase. In the simulation, it is assumed that the shifted pixel numbers along the vertical and horizontal directions introduced by the DAP and WP are 11 and 14, respectively. Finally, the hyperspectral images with 25 wavelengths could be reconstructed. The spectral resolutions of different wavelengths are not equal due to the nonlinear dispersion of the prisms. The spectral resolutions of the wavebands in the short wavelength range are higher than that in the long. The simulated measurement is shown in Fig. 5(b). It is can be seen that the LCP and RCP components in the measured image are separated with a black gap. The gap width was given as 3 pixels.

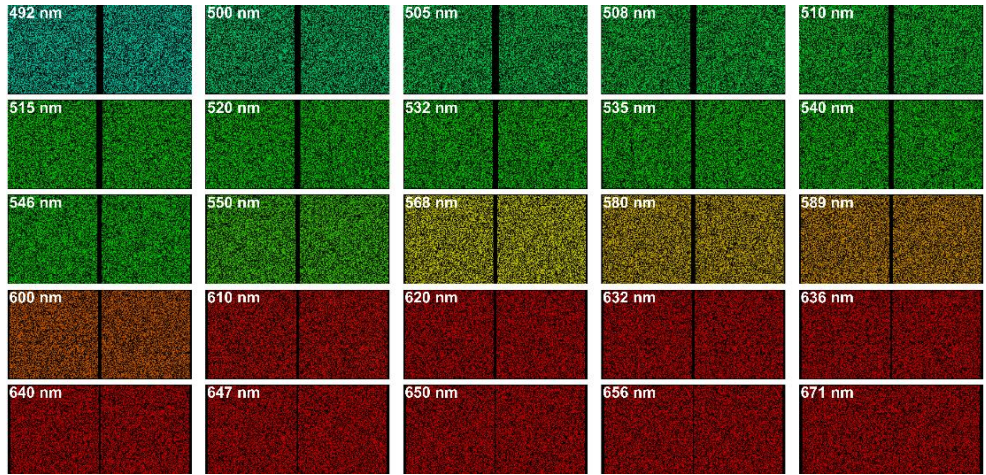

(a)

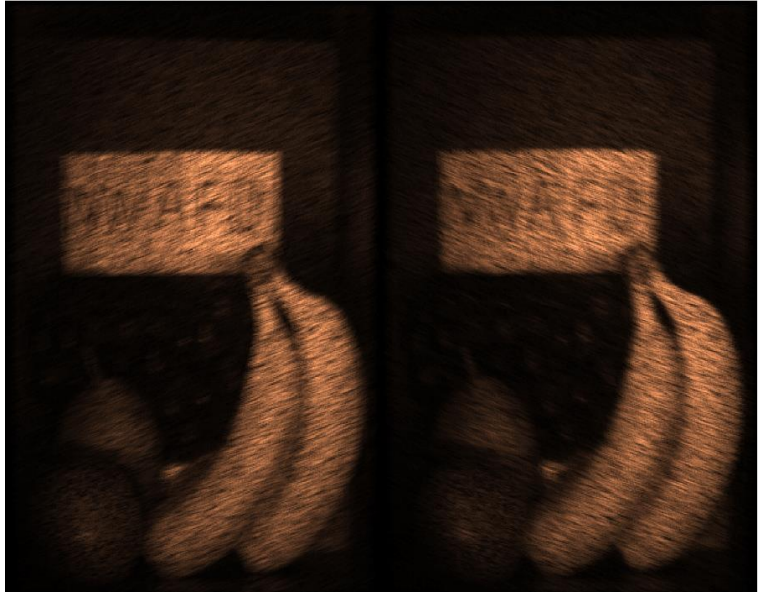

(b)

Fig. 5. Simulation results of the proposed system: (a) Simulated CAs at 25 monochromatic wavelengths; (b) Simulated measurement.

### 3.2. RECONSTRUCTION, EVALUATION, AND POLARIZATION REPRESENTATION

The hyperspectral datacube could be recovered by the TwIST algorithm. For the hyperspectral image processing, the reconstruction performance is usually estimated in two aspects, the image and spectral profile. In the follow, the reconstruction evaluation was discussed from the image and spectral evaluations, respectively.

**(a) RECONSTRUCTION AND IMAGE EVALUATION**

There were many performance evaluation criterions for reconstruction image estimation were developed by many researchers [27]. Mean-square-error (MSE), peak signal-to-noise ratio (PSNR), and structural similarity (SSIM) are adopted as three criteria to measure the reconstruction error from the perspective of image recovery in this paper. Their definitions are

$$MSE(l) = \frac{1}{M \cdot N} \sum_{i=1}^{M} \sum_{j=1}^{N} \left( \hat{f}(i,j,l) - f_0(i,j,l) \right)^2$$

$$PSNR(l) = 10\log_{10}\left( \frac{1}{MSE(l)} \right) \quad , \quad (20)$$

$$SSIM\left(\hat{f}, f_0\right) = \frac{\left(2\mu_{\hat{f}}\mu_{f_0} + C_1\right)\left(2\sigma_{\hat{f},f_0} + C_2\right)}{\left(\mu_{\hat{f}}^2 + \mu_{f_0}^2 + C_1\right)\left(\sigma_{\hat{f}}^2 + \sigma_{f_0}^2 + C_2\right)}$$

where $f_0$ and $\hat{f}$ denote the original and reconstructed images, respectively. $\mu$ and $\sigma$ are the expectation and standard deviation operators. $l$ is the index of the wavelength. $C_1$ and $C_2$ are two small constants used to avoid the instability when $\left(\mu_{\hat{f}}^2 + \mu_{f_0}^2\right)$ or $\left(\sigma_{\hat{f}}^2 + \sigma_{f_0}^2\right)$ approximates to zero.

The reconstructed RCP and LCP spectral images are shown in 错误!未找到引用源。(a) and 6(b). It is obvious that the typical features in different spectral images were distinguished. For instance, the red apple and green pear parts just can be seen in the wavelengths corresponding to the red and green colors, respectively. Furthermore, the letter 'A' is clear in the short wavelength range and blur in the long.

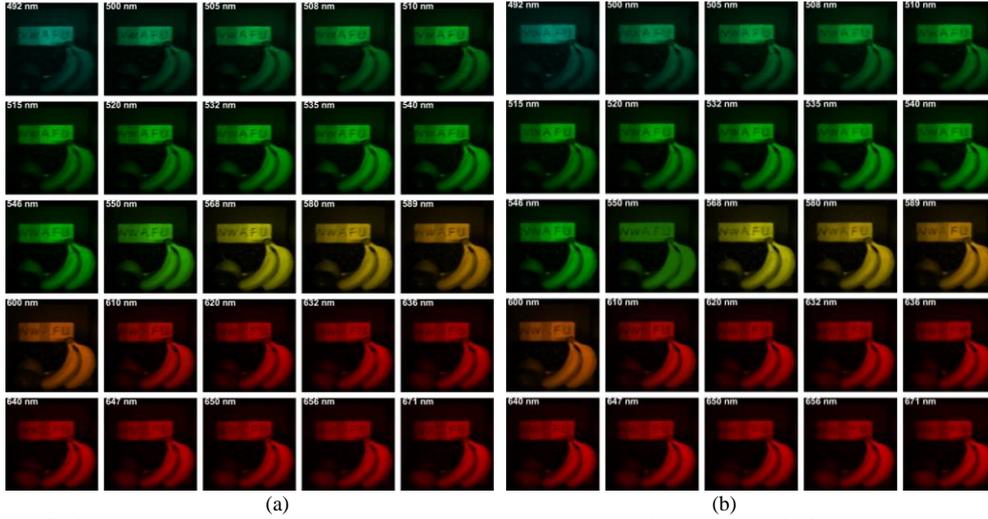

Fig. 6. Reconstructed spectral images using the TwIST algorithm: (a) Reconstructed RCP spectral images; (b) Reconstructed LCP spectral images.

To verify the robust of the proposed compressive sensing based imaging framework, different noisy hyperspectral images were generated based on the original noise free dataset in the simulation. The SNR of the contaminated images are from 10dB to 40dB with an interval of 5dB. The reconstructed hyperspectral images were evaluated by the three criteria illustrated in Eqs. (20). The criteria of different reconstructed spectral images from the simulation datasets with various noise levels are shown in Figs. 7-9, respectively.

As shown in Fig.7, the curves of MSE with respect to the RCP and LCP are depicted in Figs. 7(a) and 7(b), respectively. The MSE values depicted in the curves are all relatively small and demonstrate the fidelity of the proposed system and algorithm.

The curves for the PSNR values of RCP and LCP were calculated and shown in Fig. 8(a) and 8(b), respectively. It can be found that the minimum PSNR value is higher than 20dB for all the reconstructed images. In addition, it can be concluded that the improvement of PSNR while the SNRs of the noisy datasets higher than 30dB are not significant. Approximately, the reconstruction result could reach the optimal while the noise level of the contaminated hyperspectral images is 35dB.

As an important standard, the curves for SSIM with respect the RCP and LCP were obtained and shown in Fig. 9(a) and 9(b), respectively. It is evident that the SSIM values illustrated in the two charts are higher than 0.65. When the SNR of the original contaminated images exceeds 25dB, SSIM values are larger than 0.85. It indicates that the reconstruction is effective while the measurement not contaminates seriously. To a certain extent, the spectral image could still be reconstructed even though the measurements are featured as low SNRs.

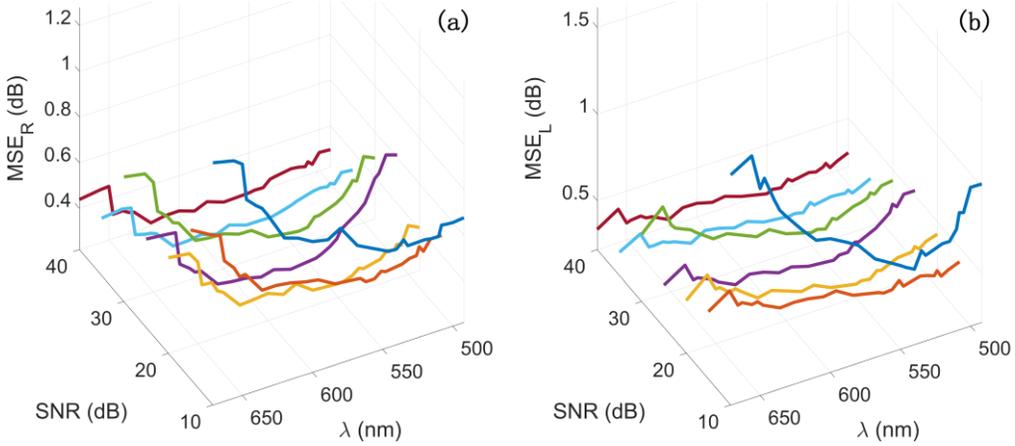

Fig. 7. MSE Vs. SNR and wavelength: (a) RCP; (b) LCP.

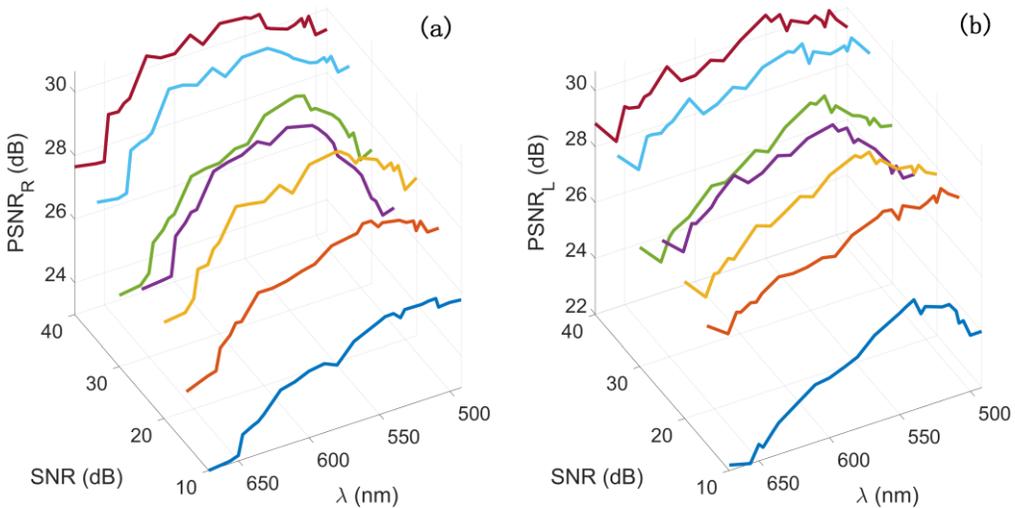

Fig. 8. PSNR Vs. SNR and wavelength: (a) RCP; (b) LCP.

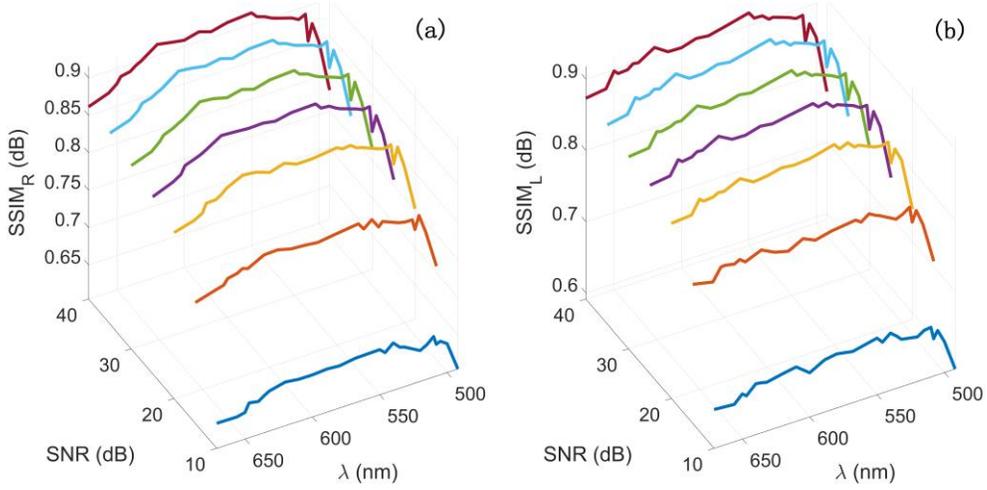

Fig. 9. SSIM Vs. SNR and wavelength: (a) RCP; (b) LCP.

**(b) RECONSTRUCTION AND SPECTRAL EVALUATION**

In the aspect of spectral recovery, Root mean square (RMS), goodness of fit coefficient (GFC), and spectral angle mapper (SAM) are three typical criteria used to evaluate the reconstruction performance [28, 29]. The definition of RMS, GFC, and SAM, respectively, are defined as

$$RMS = \frac{1}{MNL}\sum_{i=1}^{M}\sum_{j=1}^{N}\sqrt{\sum_{l=1}^{L}\left(\hat{f}(i,j,l)-f_0(i,j,l)\right)^2}$$

$$GFC = \frac{1}{MN}\left(\sum_{i=1}^{M}\sum_{j=1}^{N}\frac{\sum_{l=1}^{L}\hat{f}(i,j,l)f_0(i,j,l)}{\sqrt{\sum_{l=1}^{L}\hat{f}(i,j,l)^2}\sqrt{\sum_{l=1}^{L}f_0(i,j,l)^2}}\right). \quad (21)$$

$$\alpha = \cos^{-1}(GFC)$$

RMS represents the cumulative squared error between the original image and the reconstructed image. The smaller value of RMS demonstrates the better recovered results. Similarly, the smaller value of SAM implies the higher similarity between the recovered and original spectra. The unit of SAM is the degree. However, the larger GFC indicates the better reconstruction. The value of GFC is normalized to the range 0 to 1. RMS and SAM shown in Fig. (a) and 10(c) decrease with the increase of SNR. The minimum values are taken while SNR is 40dB. That is, the optimal reconstruction was obtained while the measurement is featured at the lowest level noise. The curves of GFC corresponding to different SNRs were plotted in Fig. (b). It is obvious that GFC climbs up to the maximum value while SNR is 40dB. The optimal result was got while the lowest noisy measurement. Thereby, the identical conclusions taken based on each evaluation criteria are consistent.

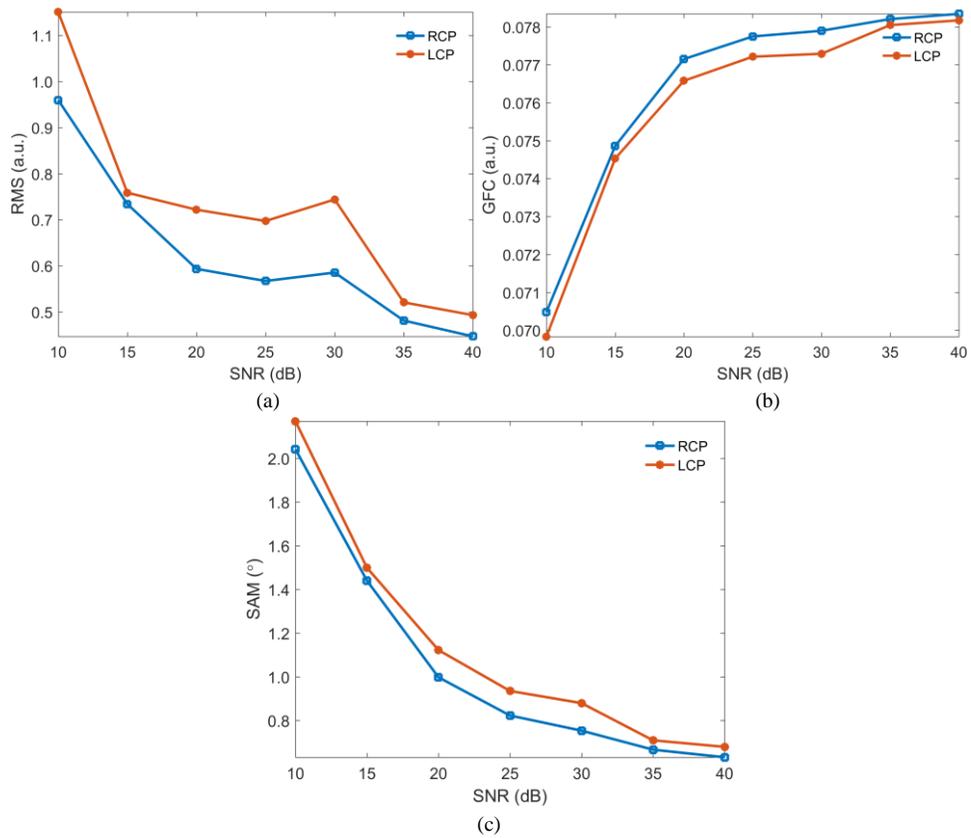

Fig. 10. Spectral evaluation results vs. SNR: RMS (a), GFC (b) and SAM (c).

Three points at different locations were marked out as shown in Fig. 4, and the spectral characteristics of the center point at each location were plotted in Fig. 10. It can be found that the reconstructed and original spectral profiles are very similar to each other. It indicates that the spectral reconstruction is effective.

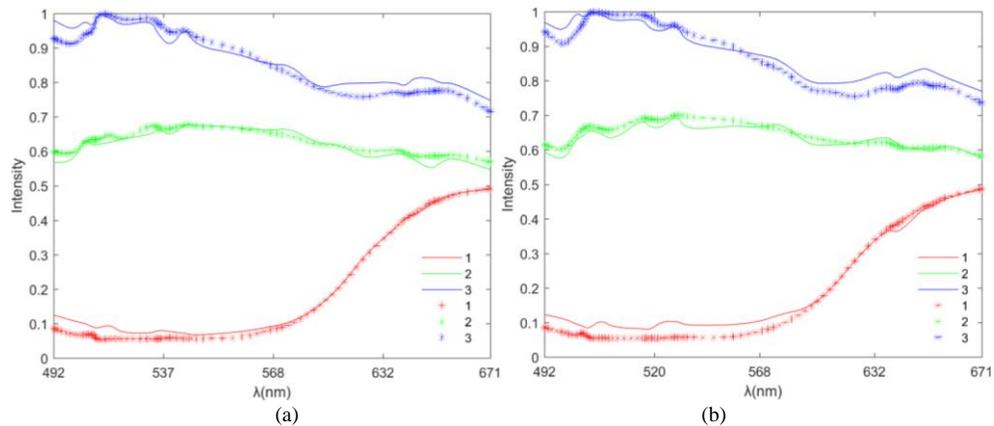

Fig. 10. Spectral profiles of three marked points as shown in Fig. 4: Spectral profiles of LCP (a) and RCP (b) .

### (c) POLARIZATION REPRESENTATION

According to the reconstructed RCP and LCP spectral images, the spectral images of the Stokes parameters $S_0$ and $S_3$ are calculated and shown in Figs. 11(a) and 11(b). Image of $S_0$ denotes a superposition of LCP and RCP, representing the total intensity of light at different bands. Image of $S_3$ is the subtraction of LCP and RCP, representing the intensity of RCP over LCP preference. It can be found that the circular polarization information is dominant in the edge of the objects.

According to Eq. (4), the spectral images of DoCP and AoCP were calculated and shown in Figs. 10(c) and 10(d). The light and shade of the image is due to the different intensity of circular polarization in different parts of the image. Hence, the shadow heavier and lighter parts in images denote more and less circularly polarized information, respectively. The surface of the fruit contains less circular polarization information. In the spectral images of AoCP and DoCP, the edge information is clearer. As shown in Figs. 10(a), the grapes with low light is hard to identify in the spectral images. However, the grapes readily can be distinguished in the AoCP and DoCP images shown in Figs. 10(c) and 10(d). It indicates that the circular polarization is a powerful tool to recognize the targets placed in the low-light environment.

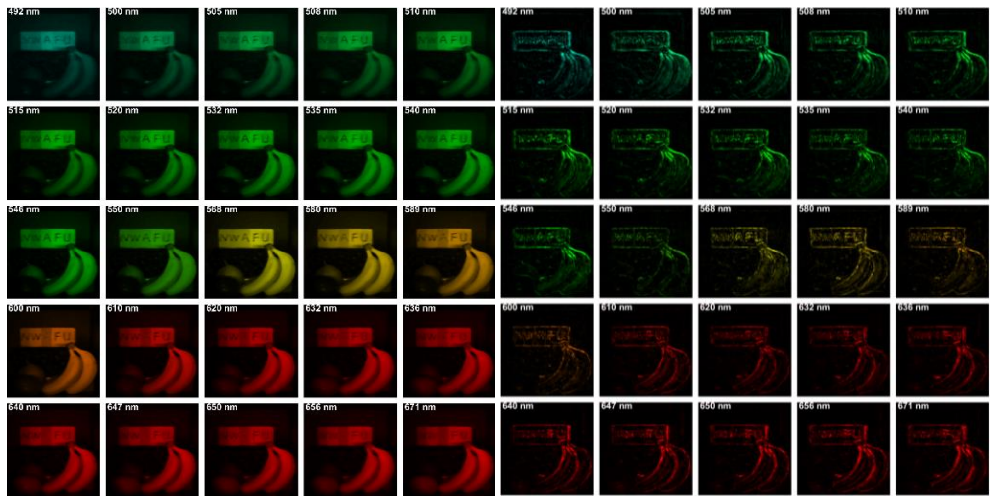

(a)          (b)

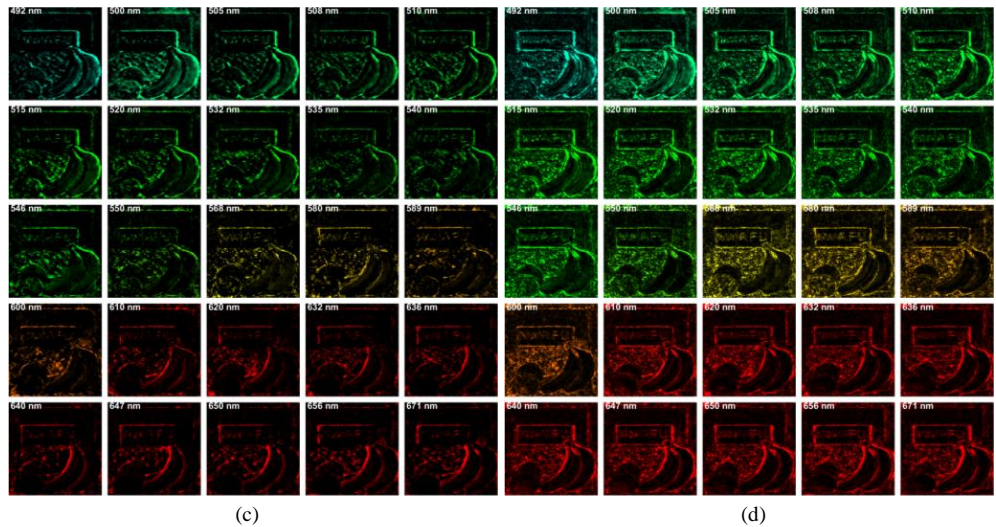

(c)　　　　　　　　　　　　　　　　　　(d)

Fig. 11. Polarization representations: (a) $S_0$; (b) $S_3$; (c) DoCP; (d) AoCP.

## 4. CONCLUSION

In summary, we propose a compressive circular polarization snapshot spectral imager based on compressive sensing paradigm. The information is encoded in the spatial domain by a coded aperture. Two dispersion prisms, DAP and WP, are used to disperse the spectral along two orthogonal directions. Meanwhile, the combination with a QWP and a WP was used as a polarization splitter to acquire the LCP and RCP components. The spectral resolution could be improved comparing with the typical CASSI with one dispersion element theoretically. The feasibility and fidelity were verified by simulation experiments. In the simulation, TwIST algorithm was used to recover the information. Our scheme provides an approach to achieve spectropolarimetry image reconstruction. It not only offers the capability of obtaining spectral and polarization information, but also relieves the pressure on data transmission, processing and storage of real-time imaging, which can be widely applied in insect classification and turbid medium imaging, and so on.

## Funding

This work was supported by the National Natural Science Foundation of China [grant numbers 11504297, 61673314, 11664004, 51979233]; Natural Science Foundation of Shaanxi Province [grant number 2019JM-302]; China Scholarship Council [grant number 201706305022]; Yangling Demonstration Zone in China [grant number 2018CXY-23]; 111 Project [grant number B12007]; Northwest A&F University [grant numbers 2452017168, 2452017166, Z109021504, Z109021508]; the 13th Five-Year Plan for Chinese National Key R&D Project [grant number 2017YFC0403203]; Fundamental Research Funds for the Central Universities [grant number 2018CX01025].

## REFERENCES

[1] J. Scott Tyo, D. L. Goldstein, D. B. Chenault, and J. A. Shaw, Review of passive imaging polarimetry for remote sensing applications, Appl. Opt. 45 (2006) 5453-5469. https://doi.org/10.1364/AO.45.005453.
[2] F. Snik, J. Craven-Jones, M. Escuti, S. Fineschi, D. Harrington, A. Martino, D. Mawet, J. Riedi, and J. Tyo, An overview of polarimetric sensing techniques and technology with applications to different research fields, Proc. SPIE 9099 (2014) 90990B. https://doi.org/10.1117/12.2053245.

[3] J. D. van der Laan, D. A. Scrymgeour, S. A. Kemme, and E. L. Dereniak, Detection range enhancement using circularly polarized light in scattering environments for infrared wavelengths, Appl. Opt. 54 (2015) 2266-2274. https://doi.org/10.1364/AO.54.002266.
[4] D. A. Miller and E. L. Dereniak, Selective polarization imager for contrast enhancements in remote scattering media, Appl. Opt. 51 (2012) 4092-4102. https://doi.org/10.1364/AO.51.004092.
[5] L. Fei, S. Xiaopeng, X. Jie, and H. Pingli, Design of a circular polarization imager for contrast enhancement in rainy conditions, Appl. Opt. 55 (2016) 9242-9249. https://doi.org/10.1364/AO.55.009242.
[6] Bénière, Arnaud, M. Alouini, F. Goudail, and D. Dolfi, Design and experimental validation of a snapshot polarization contrast imager, Appl. Opt. 48 (2009) 5764-5773. https://doi.org/10.1364/AO.48.005764.
[7] X. Meng, J. Li, T. Xu, D. Liu, and R. Zhu, High throughput full Stokes Fourier transform imaging spectropolarimetry, Opt. Express 21 (2013) 32071-32085. https://doi.org/10.1364/OE.21.032071.
[8] S. Jones, F. Iannarilli, P. Kebabian, Realization of quantitative-grade fieldable snapshot imaging spectropolarimeter, Opt. Express 12 (2004) 6559-6573. https://doi.org/10.1364/OPEX.12.006559.
[9] J. Li, B. Gao, C. Qi, J. Zhu, and X. Hou, Tests of a compact static Fourier-transform imaging spectropolarimeter, Opt. Express 22 (2014) 13014–13021. https://doi.org/10.1364/OE.22.013014.
[10] C. Zhang, Q. Li, T. Yan, T. Mu, and Y. Wei, High throughput static channeled interference imaging spectropolarimeter based on a Savart polariscope, Opt. Express 24 (2016) 23314-23332. https://doi.org/10.1364/OE.24.023314.
[11] Ali Altaqui and M. W. Kudenov, Phase-shifting interferometry-based Fourier transform channeled spectropolarimeter, Appl. Opt. 58 (2019) 1830-1840. https://doi.org/10.1364/AO.58.001830.
[12] H. Boche, R. Calderbank, G. Kutyniok, and J. Vybíral, Compressed Sensing and its Applications, Springer, Switzerland, 2015.
[13] M. F. Duarte, M. A. Davenport, D. Takhar, J. N. Laska, T. Sun, K. F. Kelly, and R. G. Baraniuk, Single-pixel imaging via compressive sampling, IEEE Signal Process. Mag. 25 (2008) 83–91. https://doi.org/10.1109/MSP.2007.914730.
[14] A. Wagadarikar, R. John, R. Willett, and D. Brady, Single disperser design for coded aperture snapshot spectral imaging, Appl. Opt. 47 (2008) B44–B51. https://doi.org/10.1364/AO.47.000B44.
[15] G. R. Arce, D. J. Brady, L. Carin, H. Arguello, and D. S. Kittle, An introduction to compressive coded aperture spectral imaging, IEEE Signal Process. Mag. 31 (2014) 105–115. https://doi.org/10.1109/MSP.2013.2278763.
[16] Tsung-Han Tsai and David J. Brady, "Coded aperture snapshot spectral polarization imaging," Appl. Opt. 52, 2153-2161 (2013). https://doi.org/10.1364/AO.52.002153.
[17] C. Fu, H. Arguello, G. R. Arce, and V. O. Lorenz, Compressive spectral polarization imaging, Proc. SPIE 9109 (2014) 91090D. https://doi.org/10.1117/12.2049870.
[18] C. Fu, H. Arguello, B. M. Sadler, and G. R. Arce, Compressive spectral polarization imaging by a pixelized polarizer and colored patterned detector, J. Opt. Soc. Am. A32 (2015) 2178–2188. https://doi.org/10.1364/JOSAA.32.002178.
[19] T.-H. Tsai, X. Yuan, D. J. Brady, Spatial light modulator based color polarization imaging, Opt. Express 23 (2015) 11912–11926. https://doi.org/10.1364/OE.23.011912.
[20] W. Ren, C. Fu, and G. R. Arce, The first result of compressed channeled imaging spectropolarimeter, Appl. Opt. 2018 JTu4A (2018) 21. https://doi.org/10.1364/3D.2018.JTu4A.21.
[21] W. Ren, C. Fu, D. Wu, Yingge Xie, and Gonzalo R. Arce, Channeled compressive imaging spectropolarimeter, Opt. Express 27 (2019) 2197-2211. https://doi.org/10.1364/OE.27.002197.
[22] M. Descour, E. Dereniak, Computed-tomography imaging spectrometer: experimental calibration and reconstruction results, Appl. Opt. 34 (1995) 4817–4826. https://doi.org/10.1364/AO.34.004817.
[23] J. M. Simon and M. C. Simon, Wollaston prism as a beam splitter in convergent light, Appl. Opt. 17 (1978) 3352-3353. https://doi.org/10.1364/AO.17.003352.
[24] C. Edward, Field Guide to Polarization, SPIE Press, Washington, 2005.
[25] G. Jing .R. Deqing, L. Chengchao, Z. Yongtian, D. Jiangpei, Z. Xi, B. Christian, Design and calibration of a high-sensitivity and high-accuracy polarimeter based on liquid crystal variable retarders, Astron. Astrophys. 17 (2017) 8. https://doi.org/10.1088/1674-4527/17/1/8.
[26] E. J. Candes, T. Tao, Near-Optimal Signal Recovery From Random Projections: Universal Encoding Strategies?, IEEE Trans. Inf. Theory 52 (2006) 5406-5425. https://doi.org/10.1109/TIT.2006.885507.
[27] Z. Wang, A. C. Bovik, H. R. Sheikh, and E. P. Simoncelli, Image quality assessment: From error visibility to structural similarity, IEEE Trans Image Process 13(2004) 600–612. https://doi.org/10.1109/TIP.2003.819861.
[28] R. Shrestha, R. Pillay, S. George, and J. Y. Hardeberg, Quality evaluation in spectral imaging – Quality factors and metrics. Journal of the International Colour Association 10(2014) 22-35.
[29] F.A. Kruse, A.B. Lefkoff, J.W. Boardman, K.B. Heidebrecht, A.T. Shapiro, P.J. Barloon and A.F.H. Goetz, The spectral image processing system (SIPS)-interactive visualization and analysis of imaging spectrometer data, Remote Sens Environ 44(1993) 145-163. https://doi.org/10.1016/0034-4257(93)90013-N.